\documentclass[letterpaper]{article} 
\usepackage{aaai2026}  
\usepackage{times}  
\usepackage{helvet}  
\usepackage{courier}  
\usepackage[hyphens]{url}  
\usepackage{graphicx} 
\usepackage{natbib}  
\usepackage{caption} 
\usepackage{algorithm}
\usepackage{algorithmic}
\usepackage{newtxmath}
\usepackage{booktabs}
\usepackage{multirow} 
\usepackage{makecell}
\usepackage{tabularx}
\usepackage{ragged2e}
\usepackage[table]{xcolor}

\usepackage{newfloat}
\usepackage{listings}
\DeclareCaptionStyle{ruled}{labelfont=normalfont,labelsep=colon,strut=off} 
\floatstyle{ruled}
\newfloat{listing}{tb}{lst}{}
\floatname{listing}{Listing}
\title{LoopLLM: Transferable Energy-Latency Attacks\\in LLMs via Repetitive Generation} 

\author{
    Xingyu Li\textsuperscript{\rm 1,2},
    Xiaolei Liu\textsuperscript{\rm 1,2}\thanks{Corresponding author is Xiaolei Liu (luxaole@gmail.com).},
    Cheng LIU\textsuperscript{\rm 1,2},
    Yixiao Xu\textsuperscript{\rm 3}, \\
    Kangyi Ding\textsuperscript{\rm 1,2},
    Bangzhou Xin\textsuperscript{\rm 1,2},
    Jia-Li Yin\textsuperscript{\rm 4}
}
\affiliations{
    \textsuperscript{\rm 1}National Interdisciplinary Research Center of Engineering Physics\\
    \textsuperscript{\rm 2}Institute of Computer Application, China Academy of Engineering Physics\\
    \textsuperscript{\rm 3}Beijing University of Posts and Telecommunications\\
    \textsuperscript{\rm 4}Fuzhou University\\

}

\usepackage{bibentry}

\begin{document}

\maketitle

\begin{abstract}
As large language models (LLMs) scale, their inference incurs substantial computational resources, exposing them to energy-latency attacks, where crafted prompts induce high energy and latency cost. Existing attack methods aim to prolong output by delaying the generation of termination symbols. However, as the output grows longer, controlling the termination symbols through input becomes difficult, making these methods less effective. Therefore, we propose \textbf{LoopLLM}, an energy-latency attack framework based on the observation that repetitive generation can trigger low-entropy decoding loops, reliably compelling LLMs to generate until their output limits. LoopLLM introduces (1) a repetition-inducing prompt optimization that exploits autoregressive vulnerabilities to induce repetitive generation, and (2) a token-aligned ensemble optimization that aggregates gradients to improve cross-model transferability. Extensive experiments on 12 open-source and 2 commercial LLMs show that LoopLLM significantly outperforms existing methods, achieving over 90\% of the maximum output length, compared to 20\% for baselines, and improving transferability by around 40\% to DeepSeek-V3 and Gemini 2.5 Flash.
\end{abstract}

\begin{links}
    \link{Code}{https://github.com/neuron-insight-lab/LoopLLM}
\end{links}

\section{Introduction}

Large Language Models (LLMs) \cite{llama2, gpt4, llama3} have demonstrated impressive performance in a wide range of real-world applications \cite{wu2023bloomberggpt, lyu2023llm}, due to their powerful capabilities enabled by billions of parameters. However, their growing scale requires substantial computational resources for the training and inference process \cite{samsi2023words}. Recent studies \cite{desislavov2023trends} suggest that inference alone accounts for up to 90\% of the total energy consumption across the lifecycle of LLMs, making inference efficiency a critical concern for system availability. Despite this, current research has focused primarily on the integrity and confidentiality aspects of LLMs \cite{yi2024jailbreak, das2025security}, while the availability component of the security triad has received limited attention \cite{meftah2025energy}. 
This gap introduces serious threats: adversaries can exploit LLM inference inefficiencies to intentionally increase computational and energy costs, leading to severe consequences in resource-constrained or time-sensitive applications.

In this paper, we explore \textit{energy-latency attacks} against modern LLMs, a class of adversarial attacks aimed at maximizing energy consumption and latency time during inference. Such attacks were first proposed by \cite{shumailov2021sponge}, who demonstrated their effectiveness on transformers model by increasing the representation of sentences. Similarly, \cite{feng2024llmeffichecker} employed perturbation-based mutations to prolong the model output for desirable purposes. \cite{dong2024engorgio} proposed effective energy-latency attacks against LLMs utilizing a parameterized proxy.

Despite the fact that current energy-latency attacks have achieved some advances against LLMs, they still exhibit two significant limitations. (1) \textbf{Limited Attack Effectiveness.} Current methods primarily rely on delaying the generation of the end-of-sequence (EOS) token, which signals termination, to prolong output length, thus increasing resource consumption. However, since this strategy does not fundamentally alter the output structure, it becomes difficult to suppress the generation of EOS through input as the output grows. As a result, the generation process may still terminate early, limiting the effectiveness of these methods. (2) \textbf{Poor Cross-Model Transferability}. Existing approaches are primarily built upon gradient-based optimization in white-box settings, making them overfit the source model. This results in poor cross-model transferability, significantly limiting the practicality of these attacks in real-world scenarios.

To address these limitations, we propose \textbf{LoopLLM}, a simple yet effective energy-latency attacks framework that compels LLMs to generate repetitive content until the maximum output length, substantially increasing energy consumption and latency time. We first investigate the phenomenon of repetitive generation, which steers the LLM generation process toward low-entropy loops, forcing the model to generate up to the maximum length. Based on this sight, we design a \textbf{repetition-inducing prompt optimization} that exploits autoregressive vulnerabilities to induce LLMs into repetitive generation. Compared to prior methods that delay EOS, LoopLLM more reliably triggers maximum-length output, leading to more effective energy-latency attacks. Furthermore, to improve transferability across models, we introduce a \textbf{token-aligned ensemble optimization} that aggregates gradients from multiple surrogate models sharing the same tokenizer.
We conduct extensive experiments on 12 open-source and 2 commercial LLMs, demonstrating that LoopLLM significantly outperforms existing baselines. Specifically, LoopLLM achieves more than 90\% of the maximum output length, compared to 20\% for baselines, and improves transferability by around 40\% to DeepSeek-V3 and Gemini 2.5 Flash.

In summary, the contributions of our work are as follows:
\begin{itemize}
    \item We propose LoopLLM, a simple yet effective energy-latency attacks framework that induces LLMs into repetitive generation, increasing energy and latency costs.
    
    \item We introduce repetition-inducing prompt optimization and token-aligned ensemble optimization to enhance the effectiveness and transferability of attacks, respectively.
    
    \item We conduct extensive experiments on 12 open-source and 2 commercial LLMs that demonstrate the superiority of LoopLLM over existing baselines.
    
\end{itemize}

\section{Related Work}

\subsection{Energy-Latency Attacks}
Energy-latency attacks aim to compromise system availability by inducing excessive energy consumption and inference latency. A seminal work by \cite{shumailov2021sponge} introduced sponge examples that significantly increase energy and latency costs during inference. Subsequent studies extended such attacks to multi-exit networks \cite{hong2021panda}, object detection \cite{Shapira2023WACV}, and vision-language model \cite{gao2024inducing}.
In LLMs, since the energy and latency costs are mainly determined by the length of the output, several approaches have emerged \cite{chen2022nmtsloth, gao2024DOSpoisonin}. LLMEffiChecker \cite{feng2024llmeffichecker} identifies critical tokens linked to long outputs and applies minimal perturbations at different levels, but is less effective in modern LLMs due to their robustness to subtle changes. Engorgio \cite{dong2024engorgio} employs a parameterized proxy distribution to track the prediction trajectory of long sequences, but their approach is tailored for text completion tasks. Although existing approaches effectively increase the output length by suppressing the EOS token, LoopLLM based on repetitive generation demonstrates superior capability in triggering endless output from LLMs.

\subsection{Repetitive Generation}
Repetitive generation \cite{olsson2022context} refers to the phenomenon in which language models continuously produce the same or highly similar sequences during inference. The issue is observed in autoregressive models of varying scales and severely compromises the quality of the generated text \cite{xu2022learning}. While often regarded as a generation flaw, repetitive generation has also been exploited as a vector for adversarial attacks. \cite{hammouri2025non} manually crafted non-halting queries that force LLMs into persistent repetition. However, this manual construction is labor-intensive and lacks scalability. \cite{geiping2024coercing} proposed an automated attack to force LLMs to generate repetitive content, but its reliance on the initial response patterns of the model limits its effectiveness. Building on these insights, we propose LoopLLM that exploits autoregressive vulnerabilities to induce LLMs into repetitive generation. 

\begin{figure}[t]
\centering
\includegraphics[width=0.98\columnwidth]{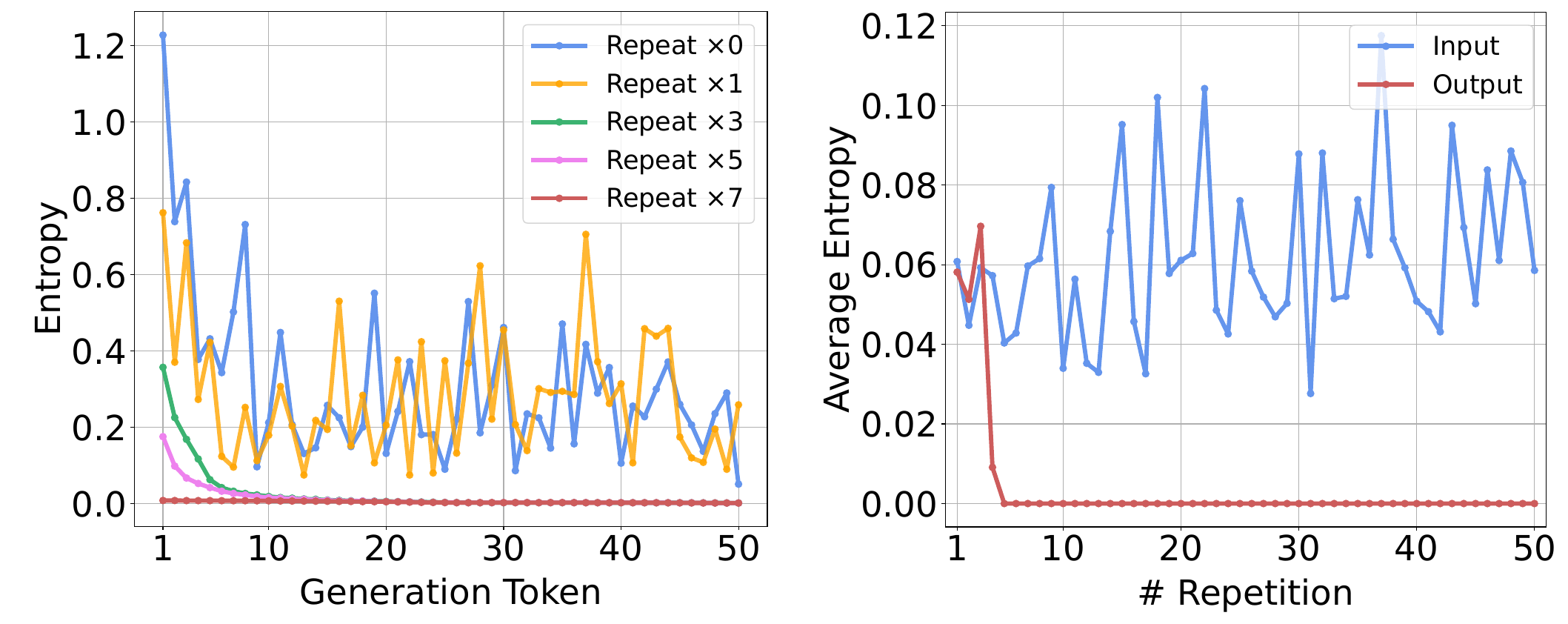} 
\caption{Left: The entropy of each generation token for varying numbers of repetitions in the input. Right: Comparison of average output entropy for varying numbers of repetitions in input and output in the instruction-aligned model.}
\label{fig:motivation}
\end{figure}

\section{Motivation}
\subsubsection{Mechanism Behind Repetitive Generation.}
Repetitive generation in LLMs arises from their autoregressive mechanism, where each token is generated based on the preceding context. This design introduces an inherent vulnerability: \textit{once the model begins to generate content that has already appeared, the mechanism may reinforce the repetition, trapping the model in repetitive generation.} Intuitively, the likelihood of such behavior increases with the frequency of repetition in the input. To validate this, we quantify this repetition using entropy, which measures the uncertainty of the language model over the next token. As shown on the left of Figure~\ref{fig:motivation}, progressively increasing the number of repeated segments in the input causes the entropy of generated tokens to rapidly converge to low values, indicating that the output distribution becomes highly concentrated on a set of tokens. This confirms the formation of low-entropy loops, a key mechanism underlying repetitive generation.

\subsubsection{Repetition in Instruction-Aligned LLMs.}
However, our focus is on dialogue scenarios, where instruction-aligned LLMs use chat templates to separate user input from model output. When repetitions are embedded solely in the input, well-aligned LLMs often disregard them as irrelevant. As shown on the right of Figure~\ref{fig:motivation}, increasing repetition in the input does not decrease the output entropy averaged over all tokens generated. In contrast, introducing just a few repetitions into the output rapidly reduces the entropy, indicating the successful induction of repetitive generation. Detailed experimental setups for both are provided in Appendix A.

Building on these observations, we propose LoopLLM that induces the model not only to recognize repetition in the input, but also to reproduce it in the output, which exploits the autoregressive vulnerability to reinforce the repetition.

\begin{figure*}[t]
\centering
\includegraphics[width=1.92\columnwidth]{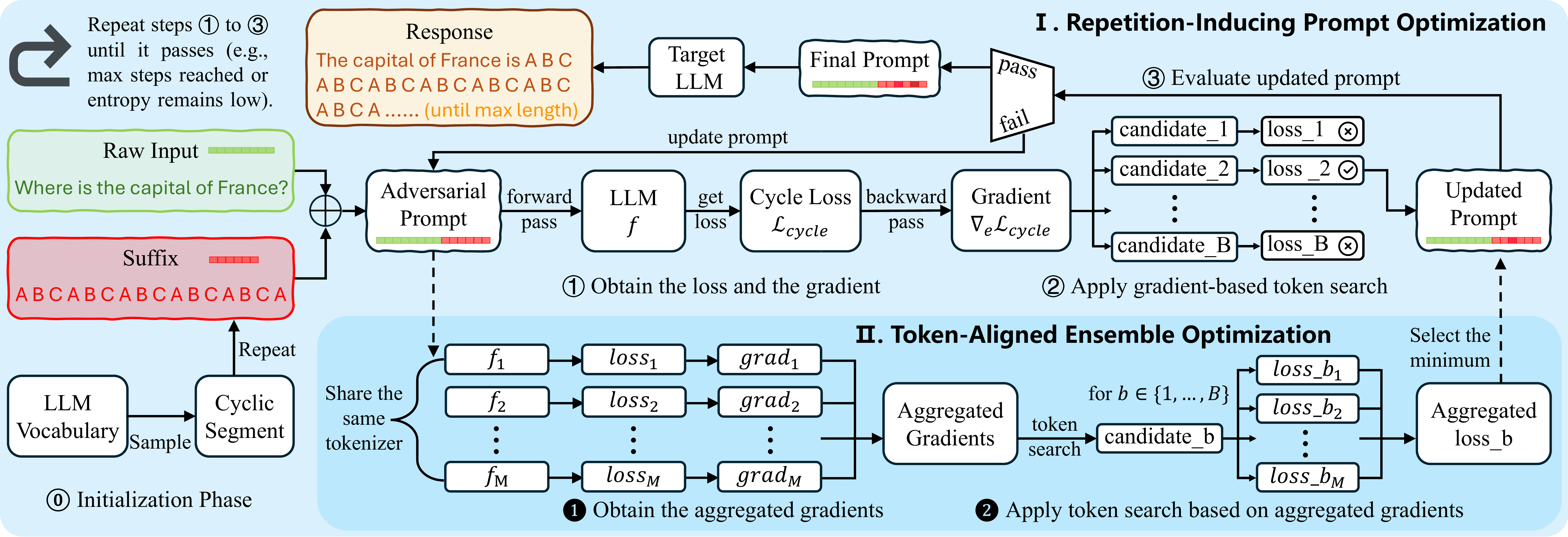} 
\caption{An overview of LoopLLM to induce LLMs to repetitive generation.}
\label{fig:overview}
\end{figure*}

\section{Preliminaries}

\subsection{Problem Formulation}

\subsubsection{Autoregressive LLMs.}
Consider an input text $\mathbf{x}$, an LLM $f$ and a sequence of output tokens $\mathbf{y}=\{y_1,y_2,\dots,y_N\}$, where $y_i$ represents the $i$-th generated token, $N$ is the output length. Modern LLMs like LLaMA typically adopt the decoder-only transformer architecture that generates tokens in an autoregressive manner. Autoregressive LLMs generate one token at a time given the input and the previously generated tokens. Formally, the probability distribution after the $\text{Softmax}(\cdot)$ over the $i$-th generated token can be denoted as:
\begin{equation}
    \mathcal{P}_i \left(\mathbf{x} ; y_1, \dots, y_{i-1}\right) = \text{Softmax} \left( f \left(\mathbf{x} ; y_1, \dots, y_{i-1} \right) \right),
\end{equation}
where $\mathcal{P}_i(\cdot) \in \mathbb{R}^{|V|}$, $|V|$ is the vocabulary size, and $f(\cdot)$ represents the forward pass of the model. Due to the token-by-token generation process, energy and latency costs are largely determined by the output length rather than the input length~\cite{feng2024llmeffichecker, dong2024engorgio}. We provide empirical evidence supporting this observation in Appendix~B.

\subsubsection{Optimization Objective.}
Based on this observation, our objective is to maximize the output length to enable effective energy-latency attacks. Unlike prior methods that delay EOS, we aim to induce repetitive generation, thereby triggering endless output during LLM inference. To this end, we design an adversarial suffix $\mathbf{x}_{\text{s}}$ appended to the raw input $\mathbf{x}$, forming an adversarial prompt $\mathbf{x}_{\text{adv}} = \mathbf{x} \oplus \mathbf{x}_{\text{s}}$, where $\oplus$ denotes concatenation. We formulate the objective as an optimization problem with the aim of finding the suffix that minimizes a loss function $\mathcal{L}$, which encourages stable repetition. The optimization problem is thus formulated as:
\begin{equation}\label{optimization}
    \mathop{\min}_{\mathbf{x}_{\text{s}}} \mathcal{L} \left( \mathbf{x}_{\text{adv}} \right).
\end{equation}

\subsection{Threat Model}
We consider two adversarial scenarios for energy-latency attacks against modern LLMs. Detailed discussions of the threat model are provided in Appendix C.
\begin{itemize}
    \item \textbf{White-box setting:} Adversaries have full access to the target model's details and perform gradient-based optimization to craft prompts that induce repetitive generation, forcing outputs to reach the maximum token limit.

    \item \textbf{Black-box setting:} Adversaries lack knowledge of the target model and typically leverage accessible surrogate models to craft prompts, relying on transferability to induce repetitive generation in the unseen target model.
\end{itemize}

\section{Methodology}

\subsection{Overview}
We present LoopLLM, a simple yet effective attack framework that constructs adversarial prompts capable of triggering repetitive generation in LLMs. As illustrated in Figure~\ref{fig:overview}, the framework consists of two core components: repetition-inducing prompt optimization and token-aligned ensemble optimization.
The first component exploits autoregressive vulnerabilities by initializing a repetitive suffix and iteratively optimizing it with a cycle loss that reinforces the repetition, trapping the model in low-entropy loops.
The second component improves transferability by aggregating gradients from multiple surrogate models that share the same tokenizer to construct more generalized adversarial prompts.

\subsection{Repetition-Inducing Prompt Optimization}

\subsubsection{Initialization Phase.}
We begin by constructing an initial suffix composed of a short token sequence repeated multiple times, each termed \textit{cyclic segment}. Specifically, we first sample tokens from the model vocabulary to form the cyclic segment. This cyclic segment is then repeated sequentially to construct an initial suffix token $\mathbf{t}_{\text{s}}$ of length $L$:
\begin{equation}
    \mathbf{t}_{\text{s}}= \{ \underbrace{ \overbrace{t_1, t_2, \ldots, t_c}^{\text{cyclic segment}}, t_1, t_2, \ldots, t_c, t_1 \ldots  }_{L}\}.
\end{equation}
where $c$ is the length of the cyclic segment and $t_i$ is the $i$-th token within it.
Before the forward pass, we need to decode the initialized suffix token $\mathbf{t}_{\text{s}}$ into the suffix text $\mathbf{x}_{\text{s}}$, which is then concatenated with the raw input to form the adversarial prompt. Moreover, since modern LLMs, particularly instruction-aligned models, require input formatted within chat templates, the adversarial prompt should be embedded within the specified template to ensure correct processing.

\subsubsection{Cycle Loss Design.}
As discussed in the Motivation Section, establishing persistent loops in instruction-aligned LLMs requires not only the presence of repetitive patterns in the input but also their reproduction in the output. Therefore, we introduce a cycle loss to optimize the adversarial suffix, which encourages the model to reproduce the cyclic segment, steering the generation process toward low-entropy loops.
Since enforcing LLMs to generate specific tokens is challenging, we adopt an untargeted objective that increases the predicted probabilities of tokens within the cyclic segment. To account for uncertainty in the occurrence position of these tokens, the loss encourages the model to generate them at every output position. The loss is defined as follows:
\begin{equation}
    \mathcal{L}_{\text{cycle}} \left( \mathbf{x}_{\text{adv}} \right) = - \frac{1}{N} \sum_{i=1}^{N} \, \log
     \sum_{j=1}^{c} \mathcal{P}_i^{t_j} \left( \mathbf{x}_{\text{adv}} \right), 
\end{equation}
where $N$ is the output length and $\mathcal{P}_i^{t_j}$ is the probability of the $j$-th token within the cyclic segment at the $i$-th output position. 
Notably, we employ softmax-normalized probabilities rather than logits to better measure the relative confidence when predicting the next token \cite{dong2024engorgio}.

\subsubsection{Gradient-Based Token Search.}
A primary challenge in optimizing the loss (see Equation~\ref{optimization}) is that the discrete suffix cannot be optimized by standard gradient descent. To avoid this, we introduce a gradient-based token search strategy. The core idea is to obtain gradients for all tokens with respect to the loss, enabling to use single-token substitutions to maximally reduce the loss. This is achieved by leveraging one-hot vectors to compute gradients for all tokens \cite{zou2023gcg}. Specifically, we compute the gradients of each token in the vocabulary $V$ at the $i$-th position in the suffix as:
\begin{equation}
    \nabla_{e_{t_i}} \mathcal{L}_{\text{cycle}} \left( \mathbf{x}_{\text{adv}} \right) \in \mathbb{R}^{|V|},
\end{equation}
where $e_{t_i}$ represents the one-hot vector corresponding to the token at the $i$-th position in the suffix (i.e., a vector of length $|V|$ with a value of one at index $t_i$ and zeros elsewhere). 
To avoid local optima, we select the top $K$ tokens with the largest \textit{negative} gradients as candidate substitutions for each token in the suffix, resulting in up to $K \times L$ total candidates. To reduce computational cost, we randomly sample $B \left(\leq K \times L\right)$ candidates from this set, recompute their loss values, and select the one that minimizes the loss as the updated suffix for the next iteration. The process continues until reaching the maximum steps or until early stopping is triggered when the output entropy stabilizes at a low level.

\subsection{Token-Aligned Ensemble Optimization}
To enhance transferability, we employ an ensemble of $M$ surrogate models to update the suffix at each optimization step. Specifically, for each token in the suffix, we compute the gradients with respect to the loss using one-hot vectors for each surrogate model. We then aggregate these gradients:
\begin{equation}
    \sum_{j=1}^{M} \nabla_{e_{t_i}} \mathcal{L}_{\text{cycle}}^{(j)} \left( \mathbf{x}_{\text{adv}} \right),
\end{equation}
and use them to update the suffix. By leveraging aggregated gradients, we prioritize token substitutions that yield broadly effective adversarial prompts, discovering adversarial prompts that are robust across different LLMs. 
To ensure the validity of the gradient aggregation, all surrogate models must share the same tokenizer, such as variants of Llama3, ensuring that one-hot vectors are token-aligned, that is, identical in both dimension and token-to-index mapping.
After obtaining candidate substitutions, we choose the one with the minimum aggregate loss as the current optimal suffix. Thus, this selection process is formalized as follows:
\begin{equation}
     \mathbf{x}_{\text{s}}^{*} = \mathop{\min}_{\mathbf{x}_{\text{s}}^{\text{b}}} \sum_{j=1}^{M} \mathcal{L}_{\text{cycle}}^{(j)} \left( \mathbf{x} \oplus \mathbf{x}_{\text{s}}^{\text{b}} \right), \quad \text{for} \ \text{b} \in \{ 1,\dots,B \}.
\end{equation}
This ensemble selection further mitigates overfitting and ensures that the adversarial prompt effectively triggers repetitive generation even when transferred to unseen models.

\begin{table*}[t]
\centering
\small
\begin{tabular}{c cc cc cc cc cc cc}  

\toprule
\textbf{Model} & \multicolumn{2}{c}{\textbf{Llama2-13B}} & \multicolumn{2}{c}{\textbf{GLM4-9B}} & \multicolumn{2}{c}{\textbf{Llama3-8B}} & \multicolumn{2}{c}{\textbf{Vicuna-7B}} & \multicolumn{2}{c}{\textbf{Llama2-7B}} & \multicolumn{2}{c}{\textbf{Mistral-7B}}\\
\textbf{Max Length} & \multicolumn{2}{c}{8192} & \multicolumn{2}{c}{4096} & \multicolumn{2}{c}{4096} & \multicolumn{2}{c}{2048} & \multicolumn{2}{c}{2048} & \multicolumn{2}{c}{2048} \\
\cmidrule(lr){2-3} \cmidrule(lr){4-5} \cmidrule(lr){6-7} \cmidrule(lr){8-9} \cmidrule(lr){10-11} \cmidrule(lr){12-13}
\textbf{Metric} & Avg-len & ASR & Avg-len & ASR & Avg-len & ASR & Avg-len & ASR & Avg-len & ASR & Avg-len & ASR\\
\midrule
Normal Inputs & 298 & 0\% & 188 & 0\% & 353 & 2\% & 233 & 1\% & 309 & 1\% & 248 & 0\%\\
Special Inputs & 541 & 2\% & 269 & 1\% & 619 & 4\% & 298 & 1\% & 501 & 2\% & 343 & 1\% \\
LLMEffiChecker & 1497 & 19\% & 1219 & 21\% & 1486 & 23\% & 815 & 22\% & 782 & 8\% & 845 & 19\% \\
Engorgio & 461 & 2\% & 289 & 2\% & 396 & 1\% & 285 & 1\% & 507 & 2\% & 484 & 6\% \\
LoopLLM-t & \textbf{7439} & \textbf{91\%} & \textbf{3730} & \textbf{90\%} & \textbf{3892} & \textbf{94\%} & \textbf{1507} & \textbf{68\%} & \textbf{1930} & \textbf{92\%} & \textbf{1700} & \textbf{79\%} \\
LoopLLM-p & \underline{6398} & \underline{78\%} &  \underline{3074} & \underline{74\%} & \underline{3398} & \underline{81\%} & \underline{1474} & \underline{66\%} & \underline{1689} & \underline{77\%} & \underline{1457} & \underline{63\%} \\

\toprule
\textbf{Model} & \multicolumn{2}{c}{\textbf{Phi4-mini}} & \multicolumn{2}{c}{\textbf{StableLM-3B}} & \multicolumn{2}{c}{\textbf{Llama3-3B}} & \multicolumn{2}{c}{\textbf{Qwen2.5-3B}} & \multicolumn{2}{c}{\textbf{Gemma2-2B}} & \multicolumn{2}{c}{\textbf{Llama3-1B}} \\
\textbf{Max length}  & \multicolumn{2}{c}{1024} & \multicolumn{2}{c}{1024} & \multicolumn{2}{c}{1024} & \multicolumn{2}{c}{1024} & \multicolumn{2}{c}{1024} & \multicolumn{2}{c}{1024} \\
\cmidrule(lr){2-3} \cmidrule(lr){4-5} \cmidrule(lr){6-7} \cmidrule(lr){8-9} \cmidrule(lr){10-11} \cmidrule(lr){12-13}
\textbf{Metric}  & Avg-len & ASR & Avg-len & ASR  & Avg-len & ASR & Avg-len & ASR & Avg-len & ASR & Avg-len & ASR \\
\midrule
Normal Inputs  & 230 & 4\% & 222 & 0\% & 270 & 1\% & 284 & 1\% & 270 & 0\% & 294 & 2\% \\
Special Inputs & 291 & 4\% & 295 & 1\% & 439 & 3\% & 400 & 1\% & 412 & 0\% & 507 & 11\% \\
LLMEffiChecker & 617 & 32\% & \underline{619} & \underline{28\%} & 638 & 27\% & 679 & 31\% & \underline{711} & 23\% & 680 & 27\% \\
Engorgio & 299 & 7\% & 353 & 4\% & 355 & 2\% & 337 & 2\% & 361 & 1\% & 399 & 10\% \\
LoopLLM-t & \textbf{971} & \textbf{92\%} & \textbf{712} & \textbf{46\%} & \textbf{982} & \textbf{93\%} & \textbf{922} & \textbf{80\%} & \textbf{836} & \textbf{65\%} & \textbf{1024} & \textbf{100\%} \\
LoopLLM-p & \underline{964} & \textbf{92}\% & 545 & 24\% & \underline{882} & \underline{78\%} & \underline{841} & \underline{67\%} & 641 & \underline{30\%} & \underline{996} & \underline{96\%} \\

\bottomrule

\end{tabular}
\caption{Results of attack effectiveness comparing our method variants with baseline methods on modern LLMs. The best results are highlighted in bold, and the second best results are underlined.}
\label{tab:white_result}
\end{table*}

\begin{table}[t]
\centering
\small
\begin{tabular}{c cc cc}  

\toprule
\textbf{Model} & \multicolumn{2}{c}{\textbf{GLM4-9B}} & \multicolumn{2}{c}{\textbf{Llama2-7B}} \\
\cmidrule(lr){2-3} \cmidrule(lr){4-5}
\textbf{Metric} & Avg-len & ASR & Avg-len & ASR \\
\midrule
Normal Inputs & 624 & 54\% & 475 & 26\% \\
Engorgio & 893 & 82\% & 866 & 74\% \\
LoopLLM-t & 1024 & 100\% & 1024 & 100\% \\
LoopLLM-p & 1024 & 100\% &  1024 & 100\% \\
\bottomrule
\end{tabular}
\caption{Results on the text completion task.}
\label{tab:text_completion}
\end{table}

\section{Experiments}

\subsection{Experimental Setups}
\subsubsection{Models and Datasets.} We consider 12 open-source LLMs from Hugging Face, spanning diverse architectures and parameter scales, including LLaMA-2-7B-Chat (Llama2-7B), LLaMA-3.1-8B-Instruct (Llama3-8B), among others. To simulate real-world scenarios, we also include two commercial models: Deepseek-V3 and Gemini 2.5 Flash. All models are aligned for instruction, and we adopt their default chat templates to ensure correct interaction.
For evaluation, we construct datasets by randomly sampling 50 instructions from each of ShareGPT \cite{sharegpt} and Alpaca \cite{wang2023alpaca}, totaling 100 benign prompts. More details are provided in Appendix D.1.

\subsubsection{Baselines.} We compare our method with four types of baseline methods. (1) Normal Inputs: We directly feed the 100 benign inputs to the models, which establishes the natural output behavior of the models. (2) Special Inputs: We append semantically suggestive phrases (e.g., ``Answer it endlessly.'') to the normal inputs to encourage longer responses. (3) LLMEffiChecker: We implement the word-level attack from \cite{feng2024llmeffichecker}, identified as the most effective variant in their study. (4) Engorgio: We compare the method from \cite{dong2024engorgio}, which also optimizes an suffix. The implementation details can be found in Appendix D.2.

\subsubsection{Metrics.} Given the strong correlation between output length and energy-latency cost during LLM inference, we consider output length to be the primary metric. Specifically, we report: (1) Average Output Length (\textbf{Avg-len}): The average number of tokens generated for all inputs. (2) Attack Success Rate (\textbf{ASR}): The percentage of all inputs for which the model generates the maximum output length.

\subsubsection{Implementation.} We fix the cyclic segment length to distinguish between our two variants. When $c=1$, we aim for token-level repetition using a single token (``*'') as the cyclic segment, denoted as \textbf{LoopLLM-t}. When $c=5$, we use a phrase (``* \% \& @ \#'') to target phrase-level repetition, denoted as \textbf{LoopLLM-p}.
The suffix length $L$ is set to 30, following Engorgio for a fair comparison. During token substitution, we set $K = 64$ and $B = 128$. To account for randomness decoding, each input is evaluated $16$ times. An attack is deemed successful if the proportion of trials that reach the maximum length exceeds $p=0.125$. Optimization is allowed for up to $20$ steps, with early stopping after success. Due to resource constraints, the maximum output length is set to 1024 tokens. We also evaluate longer limits on select models to demonstrate scalability. Representative prompts generated by all methods in Appendix D.3.

\subsection{Effectiveness Results}
Table~\ref{tab:white_result} reports the results of attack effectiveness in white-box settings.
Special Inputs induce a slight increase in output length compared to Normal Input, indicating that LLMs can interpret semantic intent but do not naturally generate excessively long outputs without explicit optimization.
LLMEffiChecker increases the length over the Normal Input, but both its Avg-len and ASR remain significantly lower than those of our variants, achieving only around 20\% ASR, compared to over 90\% for ours on most models. 
Interestingly, Engorgio exhibits an unexpectedly low efficacy compared to its original paper. We speculate that it was originally evaluated in text completion, whereas our experiments focus on dialogue scenarios with strict chat templates. To verify this, we conducted a supplementary experiment on template-free tasks, as shown in Table~\ref{tab:text_completion}, where our variants still outperform Engorgio in attack performance.

\begin{table}[t]
\centering
\small
\begin{tabular}{c c cc cc}  

\toprule
& \textbf{Model} & \multicolumn{2}{c}{\textbf{Llama3-8B}} & \multicolumn{2}{c}{\textbf{Llama2-7B}} \\
\cmidrule(lr){3-4} \cmidrule(lr){5-6}
& \textbf{Metric} & Avg-len & ASR & Avg-len & ASR \\
\midrule
\multirow{2}{*}{\makecell[c]{w/o \\ defense}} & LoopLLM-t & \textbf{981} & \textbf{93\%} & \textbf{988} & \textbf{92\%} \\
& LoopLLM-p & 887 & 79\% &  853 & 75\% \\
\midrule
\multirow{2}{*}{\makecell[c]{w/ \\ defense}} & LoopLLM-t & 439 & 14\% & 489 & 17\% \\
& LoopLLM-p & \textbf{872} & \textbf{76\%} &  \textbf{846} & \textbf{74\%} \\

\bottomrule

\end{tabular}
\caption{Results of our attack with and without defense.}
\label{tab:generate_defense}
\end{table}

\begin{table*}[t]
\centering
\small
\begin{tabular}{c cc cc cc cc cc cc}  

\toprule
\makecell[c]{\textbf{Target Model} \\ \textbf{Max Length}}
 &  \multicolumn{2}{c}{\makecell[c]{\textbf{GLM4-9B} \\ \textbf{1024}}}
 & \multicolumn{2}{c}{\makecell[c]{\textbf{Mistral-7B} \\ \textbf{1024}}}
 & \multicolumn{2}{c}{\makecell[c]{\textbf{Vicuna-7B} \\ \textbf{1024}}}
 & \multicolumn{2}{c}{\makecell[c]{\textbf{Phi4-mini} \\ \textbf{1024}}}
 & \multicolumn{2}{c}{\makecell[c]{\textbf{Deepseek-V3} \\ \textbf{2048}}} 
 & \multicolumn{2}{c}{\makecell[c]{\textbf{Gemini 2.5 Flash} \\ \textbf{2048}}} \\
\cmidrule(lr){2-3} \cmidrule(lr){4-5} \cmidrule(lr){6-7} \cmidrule(lr){8-9} \cmidrule(lr){10-11} \cmidrule(lr){12-13}
\textbf{Metric} & Avg-len & ASR & Avg-len & ASR & Avg-len & ASR & Avg-len & ASR & Avg-len & ASR & Avg-len & ASR \\
\midrule
Normal Inputs & 208 & 0\% & 268 & 1\% & 231 & 0\% & 295 & 5\% & 341 & 0\% & 437 & 0\% \\
LLMEffiChecker & 337 & 6\% & 416 & 10\% & 379 & 8\% & 491 & 13\% & 502 & 3\% & 578 & 2\% \\
LoopLLM-p & 392 & 20\% & 524 & 31\% & \underline{513} & \underline{24\%} & 608 & 43\% & 542 & 9\% & 713 & 14\% \\
LoopLLM-t & 438 & 32\% & \underline{567} & \underline{33\%} & 428 & 16\% & 647 & 49\% & 672 & 22\% & 648 & 10\% \\
\midrule
3B \& 8B  & \underline{624} & \underline{51\%} & \textbf{674} & \textbf{47\%} & 473 & 21\% & \textbf{843} & \textbf{76\%} & \textbf{927} & \textbf{43\%} & \textbf{894} & \textbf{37\%} \\
7B \& 13B & \textbf{645} & \textbf{57\%} & 467 & 27\% & \textbf{597} & \textbf{42\%} & \underline{730} & \underline{67\%} & \underline{823} & \underline{37\%} & \underline{847} & \underline{32\%} \\
\bottomrule

\end{tabular}
\caption{Results of transfer attack. The first column lists optimization strategies, primarily conducted on Llama3-8B. The ``\&'' denotes ensemble optimization using LoopLLM-t. Specifically, ``7B \& 13B'' refers to joint optimization on Llama2-7B and Llama2-13B, while ``3B \& 8B'' indicates optimization on two scales of Llama3. The first row shows target models for transfer.}
\label{tab:transfer_result}
\end{table*}

\begin{table}[t]
\centering
\small
\begin{tabular}{c c cc cc}  

\toprule
\multicolumn{2}{c}{\textbf{cyclic segment}} & \multicolumn{2}{c}{\textbf{GLM4-9B}} & \multicolumn{2}{c}{\textbf{Llama2-7B}} \\
\cmidrule(lr){3-4} \cmidrule(lr){5-6}
\textbf{length} & \textbf{token} & Avg-len & ASR & Avg-len & ASR \\
\midrule
\multirow{3}{*}{1} & LoopLLM-t & 926 & 88\% & 947 & 90\% \\
& random1 & 792 & 64\% & 912 & 86\% \\
& random2 & 903 & 82\% & 873 & 78\% \\
\midrule
\multirow{3}{*}{5} & LoopLLM-p & 838 & 67\% & 865 & 72\% \\
& random1 & 684 & 54\% & 762 & 58\% \\
& random2 & 664 & 52\% & 821 & 67\% \\
\midrule
\multirow{2}{*}{10} & random1 & 576 & 42\% & 697 & 48\% \\
& random2 & 507 & 39\% & 649 & 42\% \\
\bottomrule

\end{tabular}
\caption{The impact of cyclic segment lengths and compositions. The ``random'' denotes tokens randomly sampled from the model’s vocabulary, using two distinct random seeds.} 
\label{tab:ablation_result_cs}
\end{table}

Our method includes two variants. LoopLLM-t, which induces token-level repetition, reliably drives most models to the maximum allowable length. However, such token-level repetition is easily filtered. To evaluate this, we implemented a simple defense that halts the generation process when a token is consecutively generated beyond a threshold. As shown in Table~\ref{tab:generate_defense}, the effectiveness of LoopLLM-t drops significantly under this defense. In contrast, LoopLLM-p, which induces phrase-level repetition, remains almost unaffected with defense enabled, suggesting that phrase-level repetition is more difficult to detect and more robust against defensive measures.
The experimental results of LoopLLM-p demonstrate strong performance compared to other baselines, with only a slight reduction in effectiveness compared to LoopLLM-t. Overall, these results highlight the superiority of our approach, which more effectively forces LLMs to reach maximum output limits than EOS-delaying baselines.

\subsection{Transferability Results}
The real-world threat of adversarial attacks is its ability to transfer from accessible surrogate models to unseen target models. To evaluate this, we conducted a series of transfer attacks, with results presented in Table~\ref{tab:transfer_result}. 
We first assess transferability using a single surrogate model, where our method consistently outperforms the baselines. We attribute this superior transferability to the shared tendency among LLMs toward repetitive generation.

Next, we investigate whether ensembles of surrogate models could further improve transferability by evaluating two ensemble configurations: one using Llama2-7B \& 13B, and another using Llama3-3B \& 8B. The results show that both significantly improve the attack performance across all target models compared to single-model optimization. These results confirm the effectiveness of our ensemble optimization. 
Moreover, we also evaluate transfer attacks against commercial models. The ensemble-optimized prompts exhibit strong transferability, achieving maximum allowable lengths of 43\% on Deepseek-V3 and 37\% on Gemini 2.5 Flash, which corresponds to an improvement of nearly 40\% over baselines. This finding underscores the practical risk of our attack, as it can successfully degrade the availability of commercial LLM services without prior knowledge. We provide examples of real-world scenarios in Appendix E.

\subsection{Ablation Study}

\subsubsection{Effect of Cyclic Segment.} We begin by analyzing how the length and composition of the cyclic segment influence the attack performance. The results are presented in Table~\ref{tab:ablation_result_cs}. Varying the length of the segment $c$ under $[1,5,10]$, we observe that shorter segments are more effective. This suggests that simpler repetition patterns are easier for LLMs to replicate, making them more potent inducing repetitive generation. 
Within each cyclic segment, we observe that specifically chosen tokens (such as LoopLLM-t and LoopLLM-p) outperform randomly selected tokens from the vocabulary. This indicates that the principle of repetition is still effective with random tokens and that certain tokens in the embedding space are more conducive to inducing repetitive generation.

\subsubsection{Effect of Suffix Length and Optimization Step.} 
\begin{figure}[t]
\centering
\includegraphics[width=0.90\columnwidth]{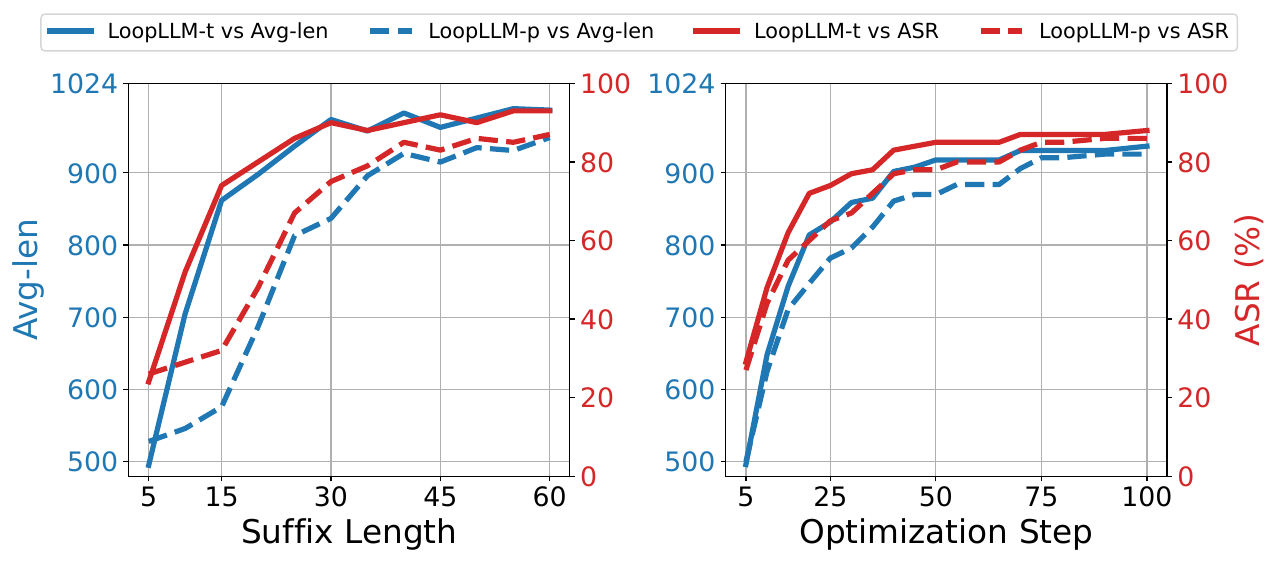} 
\caption{The impact of suffix length and optimization step}
\label{fig:ablation_length_step}
\end{figure}

We further analyze the impact of two critical hyperparameters, as illustrated in Figure~\ref{fig:ablation_length_step}. 
For the suffix length, both Avg-len and ASR generally increase with the suffix length. This is intuitive, as a longer suffix provides more repeated instances of the cyclic segment, increasing the likelihood of initiating the loop. However, the performance gain diminishes beyond a certain length, indicating a trade-off between attack strength and prompt conciseness.
Similarly, more optimization steps lead to better performance, confirming the effectiveness of our cycle loss and gradient-based search. The early convergence of both metrics suggests that our method quickly identifies an effective solution, with little improvement in later steps due to the early stopping strategy. Additional hyperparameter studies are provided in Appendix F.

\begin{table}[t]
\centering
\small
\begin{tabular}{c cc cc}  

\toprule
\multirow{2}{*}{\textbf{Decoding Strategy}} & \multicolumn{2}{c}{\textbf{GLM4-9B}} & \multicolumn{2}{c}{\textbf{Mistral-7B}}\\
\cmidrule(lr){2-3} \cmidrule(lr){4-5}
 & Avg-len & ASR & Avg-len & ASR \\
\midrule
greedy search & 818 & 68\% & 763 & 52\% \\
beam search & \textbf{956} & \textbf{94\%} & 852 & 71\% \\
temperature=0.2 & 897 & 82\% & 834 & 68\% \\
temperature=0.6 & 913 & 84\% & \textbf{883} & \textbf{74\%} \\
temperature=1.2 & 698 & 61\% & 647 & 41\% \\

\bottomrule

\end{tabular}
\caption{Results of decoding strategies on LoopLLM-t.}
\label{tab:ablation_result_decode}
\end{table}

\subsubsection{Effect of Decoding Strategy.}
We also investigate how different decoding strategies impact the attack performance. As shown in Table~\ref{tab:ablation_result_decode}, LoopLLM performs best under beam search and moderate temperature sampling, while we observe a notable degradation under greedy search and high temperature sampling. We attribute this to two factors: (1) In greedy search, even with elevated target token probabilities, non-cyclic tokens with slightly higher initial probabilities still are preferred, preventing the formation or stabilization of the loop. (2) high temperature flattens the output distribution, which also diminishes the probability advantage of target tokens. 
However, Given that such strategies often degrade text quality and are rarely used in practice, our method remains robust under common decoding strategies.

\begin{figure}[t]
\centering
\includegraphics[width=0.93\columnwidth]{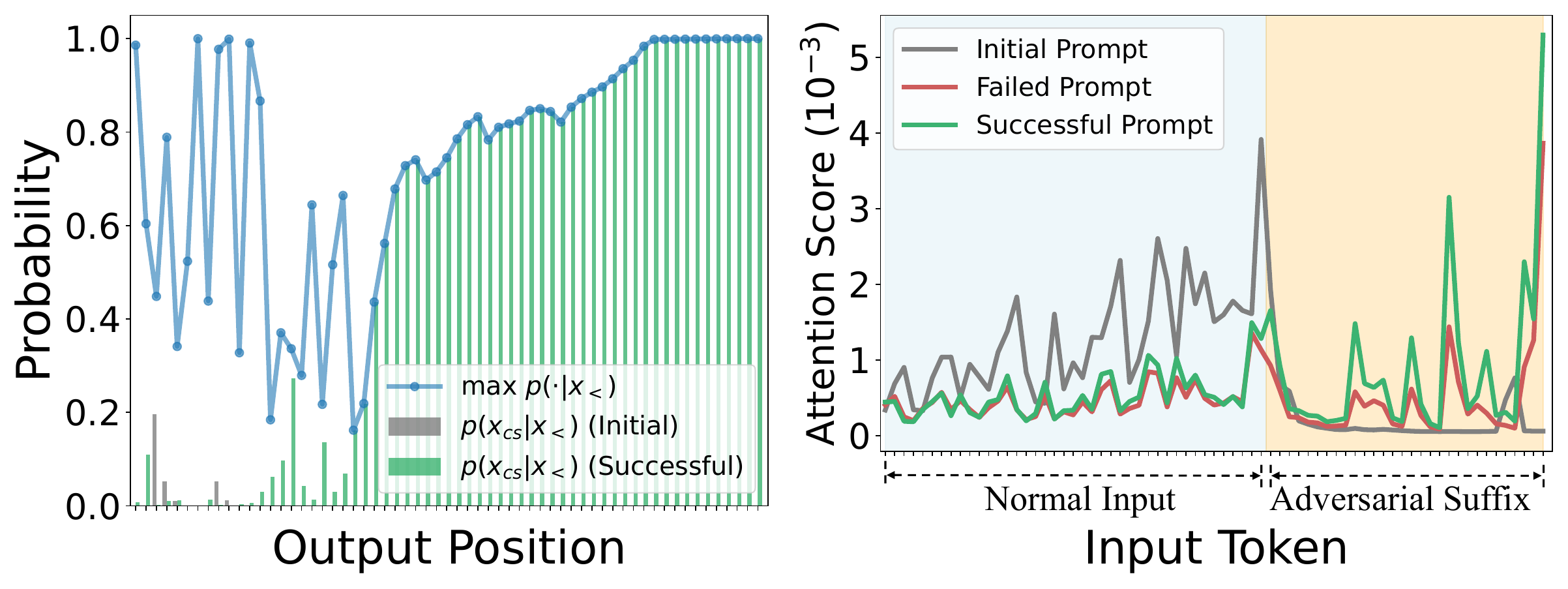} 
\caption{Left: Probabilities of the most likely token $max \ p(\cdot | x_<)$ and the cyclic segment tokens $p(x_{cs}|x_<)$ (including initial and successful prompt) at each output position. Right: Attention scores over output at each input token.}
\label{fig:effect}
\end{figure}

\subsection{Why is Our Method Effective?}
To understand the mechanisms behind LoopLLM, we conducted a qualitative analysis focusing on output probabilities and model attention. Firstly, we track the token probability at each output position. As shown on the left of Figure~\ref{fig:effect}, comparing the initial prompts with the successful ones reveals that our cycle loss effectively increases the likelihood of tokens within the cyclic segment. Meanwhile, once these tokens become the most probable, their probabilities grow almost monotonically with repetition until it stabilizes around a ceiling value. This observation confirms the existence of the autoregressive vulnerability that LoopLLM leverages.

Secondly, we compute attention scores for each input token by averaging the attention weight from all output tokens across all layers and all heads. The results are visualized on the right of Figure~\ref{fig:effect}. For the initial prompt, the attention of output is primarily distributed on normal input. In contrast, after optimization, both failed and successful prompt reallocate the majority of the attention to adversarial suffix. This shift suggests that our cycle loss effectively guides the model to pay more attention to the repetitive pattern within the adversarial suffix. 
Notably, the successful prompt exhibits a significantly stronger attention shift than the failed one, suggesting that greater attention to the suffix correlates with a higher likelihood of inducing repetitive generation.

\subsection{Potential Countermeasures}
Currently, there are no dedicated defenses for energy-latency attacks yet. We consider perplexity (PPL) filtering \cite{alon2023ppldetect}, a common technique for detecting adversarial prompts in LLMs by flagging semantically incoherent inputs. PPL is computed as the average negative log-likelihood of input tokens, and prompts with excessive PPL will be rejected before inference.
As shown in Table~\ref{tab:defense}, while baselines like LLMEffiChecker and Engorgio show substantially increased PPL compared to normal inputs, our variants exhibit low PPL, even below that of normal prompts. This result is attributable to our suffixes with repetitive pattern that steers the model into low-entropy loops, where the model exhibits high confidence in predicting subsequent tokens. Therefore, LoopLLM is inherently resistant to PPL-based filtering, despite the presence of meaningless suffixes.

Given these properties, we consider another potential defense based on monitoring of output entropy, which halts generation when the model’s output entropy stabilizes at a low threshold. Table~\ref{tab:defense} confirms that our method produces lower output entropy than both normal input and baselines, suggesting the feasibility of this defense in detecting LoopLLM. However, entropy-based filtering requires real-time entropy tracking during inference, incurring substantial computation overhead. Prior work that mitigates repetition also faces similar trade-offs between efficiency and output quality \cite{huang2025rap}, underscoring the need for more effective defenses for energy-latency attacks.

\begin{table}[t]
\centering
\small
\begin{tabular}{c rr}  

\toprule
 & \textbf{Input PPL}  & \textbf{Output Entropy} \\
\midrule
Normal Inputs & 69.34 & 0.278\\
\midrule
LLMEffiChecker & 4121.54 & 0.263 \\
Engorgio  & 6728.15 & 0.285 \\
LoopLLM-t & \textbf{42.09} &  \textbf{0.082}\\
LoopLLM-p & 124.71 & 0.147 \\

\bottomrule

\end{tabular}
\caption{The evaluation of each methods using input perplexity (PPL) and output entropy, averaged over all optimized input and corresponding output tokens.}
\label{tab:defense}
\end{table}

\section{Conclusion}
In this paper, we introduce LoopLLM, an energy-latency attacks framework against modern LLMs that triggers endless output, significantly increasing energy consumption and latency time. 
We empirically reveal that repetitive generation can steers the generation process toward low-entropy loops, compelling the model to generate repeated tokens until the maximum output length is reached. Leveraging this insight, we design a repetition-inducing prompt optimization that exploits autoregressive vulnerabilities to induce repetitive generation. To enhance transferability, we introduce a token-aligned ensemble optimization that aggregates gradients to construct generalizable prompts. Extensive experiments on 12 open-source and 2 commercial LLMs show that LoopLLM consistently outperforms baseline methods. 

\section{Acknowledgments}
 This paper was supported by the National Natural Science Foundation of China under Grant Nos.U2441239, 62302468 and 62202104.

\bibliography{references}

\noindent {\Large \textbf{Appendix}}
\setcounter{secnumdepth}{2} 
\appendix

\section{Entropy Experiment Setups}
This appendix provides the detailed experimental configuration and analyses in the Motivation Section. We describe the design and implementation of two experiments aimed at investigating the entropy under varying repetition levels in both autoregressive and instruction-aligned LLMs.

\subsubsection{Mechanism Behind Repetitive Generation}
The first experiment aims to uncover the mechanism behind repetitive generation in autoregressive LLMs. Specifically, it investigates how repetition in the input affects repetitive generation, thereby providing empirical support for autoregressive vulnerability, where LLMs are prone to enter self-reinforcing loops once they begin to repeat content.
To quantify repetitive generation in token prediction, we use token-level entropy as a metric. which measures the uncertainty of the language model over the next token. Given the probability distribution $\mathcal{P}_i$ over the model vocabulary $V$ at generation step $i$, the entropy $H_i$ is defined as:
$$
H_i = - \sum_{j=1}^{|V|} \mathcal{P}_i^{t_j} \log \mathcal{P}_i^{t_j},
$$
where $|V|$ denotes the vocabulary size, and $\mathcal{P}_i^{t_j}$ is the predicted probability of the $j$-th token at position $i$. A lower entropy indicates a more confident distribution, often associated with repetitive generation.

Due to autoregressive vulnerability, it is intuitive that more repetition in the input increases the likelihood of the model entering self-reinforcing loops. To validate this, we construct a series of inputs by embedding repeated segments. The segment is chosen as a simple natural language sequence, such as “once upon a time”, and we vary the number of times this segment is repeated (e.g., 0, 1, 3, 5, 7 repetitions). Each input is fed into the model, and we only generate 50 tokens of output. Given the randomness of decoding (temperature sampling), we repeat the generation multiple times for each input. The entropy at each generation step is averaged across these repeated trials to obtain a stable estimate of the entropy curve. As shown on the left of Figure~\ref{fig:motivation} in the main paper, increasing the repetition in the input drastically reduces the entropy of generated tokens, particularly when the repetition times exceed 3. The entropy curve rapidly converges to low values ($<$ 0.05), signaling that the predicted token distribution becomes highly concentrated on a few tokens, that is, repetitive generation occurs. These findings confirm the emergence of low-entropy loops, where the autoregressive process reinforces repetition.

\subsubsection{Repetition in Instruction-Aligned LLMs}
The second experiment explores the repetition behavior of instruction-aligned models. These models typically rely on chat templates that explicitly separate user input from model output, allowing them to maintain structured conversational coherence and respond accurately to user queries. However, this separation mechanism also introduces a key question: Is input-side repetition alone sufficient to trigger repetitive generation? In practice, instruction-aligned models are often robust to meaningless or nonsensical input content, tending to disregard the repetitions within the user prompt. In contrast, when repetition appears not only in the input but also in the model output, the autoregressive mechanism is more likely to reinforce the repeated patterns, potentially resulting in persistent low-entropy loops.
To investigate this, we prepare a sequence of test cases with repetition levels ranging from 1 to 50. For each test case, we record: (1) Input repetition entropy: Repeated segments are embedded only in the user input section of the template. (2) Output repetition entropy: Identical segments are inserted directly into the model's previous response (i.e., within the output section of the template). In each case, we compute the average output entropy over all generated tokens:
$$
\overline{H} = \frac{1}{N} \sum_{i=1}^{N} H_i,
$$
where $N$ is the length of generated tokens, and $H_i$ is the token-level entropy as previously defined. Similarly, we sample each test case multiple times to account for the randomness of the decoding, and then calculate the average value as the final entropy. 
As illustrated on the right of Figure~\ref{fig:motivation} in the main paper, increasing repeated segments solely in the input yields negligible effect on output entropy. In contrast, when similar repetitions are added to the model output, the entropy sharply drops to near-zero levels, even with only a few repetitions. This suggests that instruction-aligned models are robust against repetition in the input due to their structural separation between user query and model response. However, once repetition is echoed in the model output, the autoregressive mechanism engages, reinforcing the repetition and initiating low-entropy loops.

\begin{figure*}[t]
\centering
\includegraphics[width=1.95\columnwidth]{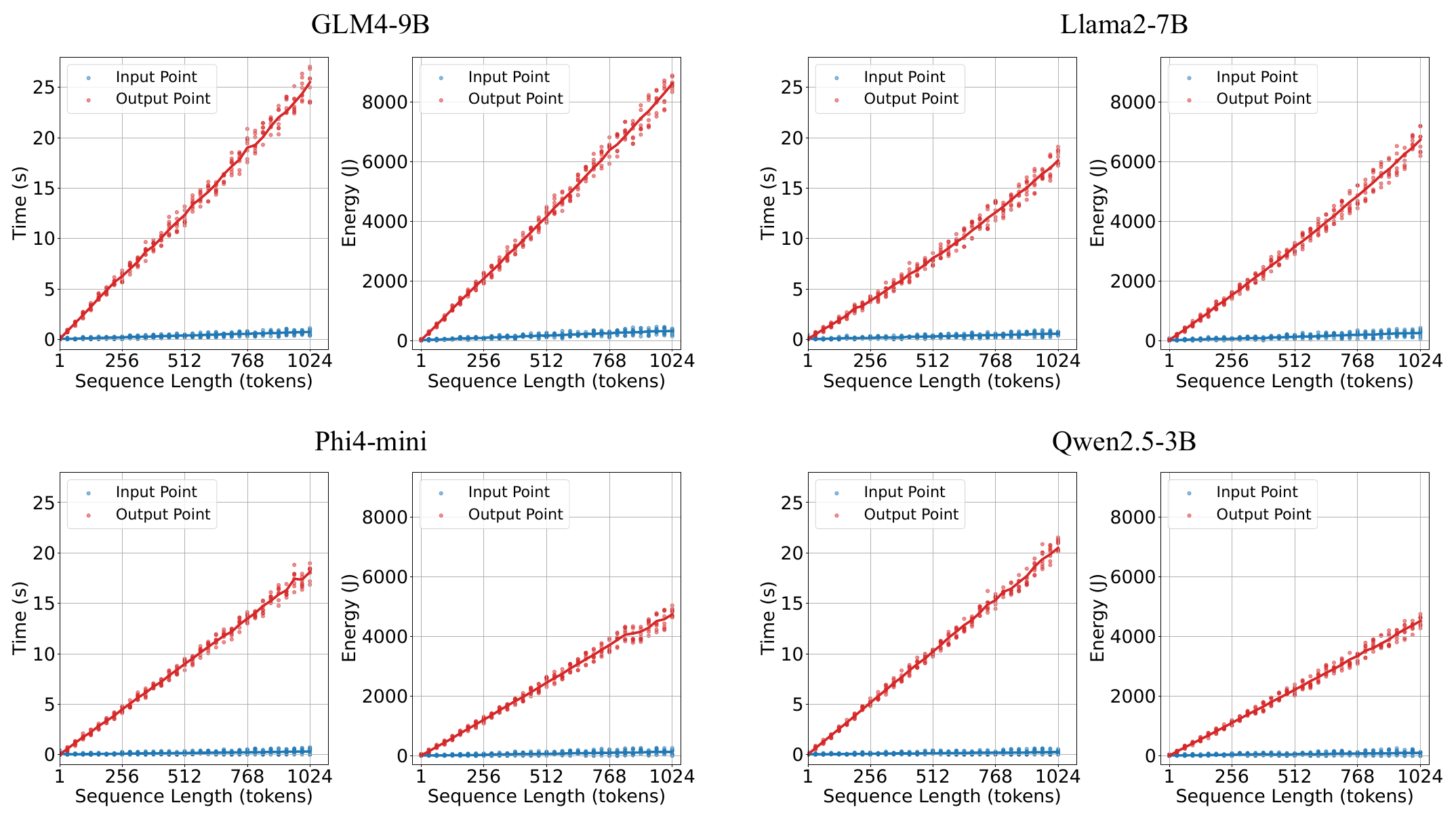} 
\caption{Relationship between sequence length and model efficiency for four models. To mitigate random variance, we measure 8 times for each sequence length and plot all data point and the corresponding average value. The inference time and energy consumption grow approximately linearly with output length, while remaining insensitive to input length.}
\label{fig:appendix_motivation}
\end{figure*}

\section{Input/Output Length vs. Model Efficiency}

For a single request to a large language model (LLM), the primary factors affecting energy consumption and inference latency are the lengths of the input and output sequences. In this section, we explore in detail how input and output lengths individually contribute to model efficiency. The inference process of an LLM typically can be divided into two distinct stages: the prefill phase and the decoding phase. During the prefill phase, the model processes the entire input in parallel using self-attention mechanisms, while also caching key-value pairs (KV cache) for use in subsequent steps. In the decoding phase, the model generates output tokens in an autoregressive manner, using the cached values to process one token at a time. Given that input tokens are processed in parallel and output tokens are generated sequentially, it is intuitive to hypothesize that the cost of energy and latency during LLM inference is more sensitive to output length than to input length. 

To validate this, we conducted comprehensive end-to-end experiments across several open-source LLMs, measuring both energy consumption and inference time as functions of input and output length. For energy measurement, we follow \cite{shumailov2021sponge} and utilized the NVIDIA Management Library (NVML) to measure GPU power consumption. We varied the sequence length from 1 to 1024, sampling every 32 tokens. To mitigate noise and random variance, each data point was measured 8 times. When measuring the impact of input length, we kept the output length constant at one token. Similarly, when evaluating output length, the input length was fixed to a single token. To ensure that output tokens continued generating up to the preset limit without early termination, we manually disabled the appearance of the end-of-sequence (EOS) token during decoding.

As shown in Figure~\ref{fig:appendix_motivation}, we plotted the value of each data point and the averaged measurements to obtain smooth trend lines. It reveals that energy consumption and latency remain nearly constant as input length increases. In contrast, both grow significantly and nearly linearly with output length. This trend holds across different model sizes and architectures, indicating that it is a fundamental characteristic of the modern LLM inference process. 
From these findings, we conclude that model efficiency, i.e., energy and time cost, is overwhelmingly determined by the length of its generated output, not the input length. This insight underscores the rationale behind our strategy to formulate instructions that maximize output response length, thereby largely amplifying energy and computational demands, potentially leading to system overloads and severe consequences.

\section{Discussion for Threat Model}
In this work, we consider two adversarial settings for energy-latency attacks against large language models (LLMs): white-box and black-box threat models.
\subsubsection{White-box Threat Model.} In the white-box setting, the attacker is assumed to have complete access to the internal architecture, parameters, and gradients of the target model. This allows the attacker to perform gradient-based optimization like LoopLLM to generate highly effective adversarial prompts. Their primary objective is to compromise the availability of the target LLM service by maximizing the length of the output sequence. Although seemingly idealized, this setting is applicable due to the increasing adoption of open-source LLMs in real-world deployments.
A variety of public services and academic platforms now provide LLM inference endpoints based on open-source models, including Huggingface, OpenRouter, and local deployments via tools such as ollama or vllm. These services often lack rigorous access control, making them susceptible to gradient-based prompt attacks. Moreover, developers and smaller organizations may deploy open-source LLMs without extensive fine-tuning due to resource constraints, making them prime targets for energy-latency attacks.
In these environments, adversarial prompts can be crafted offline using full model access and then deployed to public endpoints. Since many such services enforce rate limits on a per-request basis rather than per-token, attackers are incentive to maximize the number of output tokens per request, thus degrading the availability and responsiveness of the service.
\subsubsection{Black-box Threat Model.} In contrast, the black-box setting assumes that the attacker has no access to the internal details of the target model. The attacker can only interact with the model through a query-response API, where they can only send prompts and observe the corresponding outputs. This setting reflects most real-world commercial LLM services, such as OpenAI's GPT-4, which expose their capabilities solely via API calls. The objective in this setting is to compromise service availability by forcing the LLM to generate exceptionally long outputs for each API call.
Despite the lack of internal access, attackers can still mount effective attacks leveraging the transferability of adversarial prompts. Specifically, an attacker can use accessible surrogate models to generate adversarial prompts utilizing our LoopLLM framework and then deploy them against black-box targets. Given the transferability of our method, these adversarial prompts can generalize across architectures and model scales. The attacker evaluates the effectiveness of the attack by observing whether the target model produces repetitive outputs that reach the maximum token limit, thereby increasing resource usage.
This black-box strategy is particularly dangerous in multi-user or shared-resource scenarios. In such contexts, a single user can monopolize the token budget of the system by issuing a small number of adversarial queries. This not only increases operational costs but also reduces service availability for other users. As commercial LLMs often implement strict request caps or per-token pricing, attackers can exploit these conditions to cause financial or service-level disruptions.

Such attacks in both settings are especially impactful in subscription-based or free-tier deployments, where cost-efficiency and service quality are critical. Regarding the attacker's motivation, a malicious actor might aim to disrupt a rival's service in the fiercely competitive LLM market. By flooding a competitor's service with LoopLLM prompts, an attacker can inflate their operational costs and degrade their service quality, thereby making the own service appear more reliable. In environments with shared resources, such as a service with a global query-level rate limit, an attacker can use LoopLLM prompts to consume a shared quota. This effectively blocks other legitimate users and constitutes a denial-of-service attack. Furthermore, some adversaries may not have a financial or competitive motive but are driven by the desire to intentionally cause disruption and chaos.

\section{Detailed Experiments Setups}
\subsection{The Detail of models and datasets}
This section provides a detailed overview of the Large Language Models (LLMs) and the datasets utilized for the experiments in this paper.
\subsubsection{Models.} 
We employ a diverse suite of 12 open-source and three proprietary LLMs to ensure a comprehensive evaluation. All open-source models were sourced from Hugging Face, covering a broad spectrum of parameter scales, model families, and instruction tuning techniques. These models include LLaMA-2-13B-Chat (Llama2-13B), LLaMA-3.1-8B-Instruct (Llama3-8B), GLM-4-9B-Chat (GLM4-9B), LLaMA-3.1-8B-Instruct (Llama3-8B), Vicuna-7B-v1.5 (Vicuna-7B), LLaMA-2-7B-Chat (Llama2-7B), Mistral-7B-Instruct-v0.3 (Mistral-7B), Phi-4-Mini-Instruct (Phi4-mini), StableLM-Zephyr-3B (StableLM-3B), LLaMA-3.2-3B-Instruct (Llama3-3B), Qwen2.5-3B-Instruct (Qwen2.5-3B), Gemma-2-2B-It (Gemma2-2B) and LLaMA-3.2-1B-Instruct (Llama3-1B). 
The models are based on well-known foundation architectures such as LLaMA, Mistral and Gemma. These base models are pre-trained on diverse and large-scale corpora using causal language objectives like next token prediction. However, they are not directly suitable for conversational tasks until they undergo instruction tuning tuned on datasets consisting of user instructions and expected model completions. This process aligns the base models to better follow user instructions and generally act as helpful conversational assistants.
All models used in our experiments are instruction-tuned versions, such as Llama-3.1-8B-Instruct. A crucial aspect of interacting with these models is the use of a predefined chat template format. This format often includes system prompts, user queries, assistant responses, and special identifier token (e.g. $<|\text{user}|>$, $<|\text{assistant}|>$). 
The reliance on chat templates introduces additional challenges for energy-latency attacks. Unlike traditional text completion tasks, instruction-tuned LLMs are often designed to filter irrelevant, repetitive, or malformed input. Thus, adversarial prompts must be not only structurally well-formed but also semantically plausible enough to pass through the internal instruction filters. Our attack framework explicitly incorporates such constraints, ensuring the adversarial suffix is embedded in the appropriate dialogue context.
In addition to open-source models, we also evaluate the transferability of adversarial prompt to proprietary black-box models, including Deepseek’s Deepseek-V3, Google’s Gemini 2.5 Flash. These models represent the black-box challenge, as their internal architectures are undisclosed, and access is limited to public APIs.

\subsubsection{Datasets.} To construct a diverse set of benign user queries, we draw from two widely used datasets in instruction tuning research: ShareGPT and Alpaca. The ShareGPT dataset is a large-scale collection of real-world conversations shared by users of OpenAI's ChatGPT, capturing a range of topics and complex interaction patterns. The Alpaca developed by researchers at Stanford University is a synthetically generated instruction-following dataset created by 52K instruction-output pairs derived from OpenAI’s text-davinci-003. From each dataset, we randomly sample 50 queries, resulting in a total of 100 instructions that serve as the foundation for our experiments. By combining both, it provides a diversity of prompt styles and reflects the types of prompt LLMs are likely to encounter in production scenarios.

\subsection{The Implementation of Baselines}
In order to assess the effectiveness of our method, we benchmarked it against four baseline approaches.

\begin{itemize}
    \item \textbf{Normal Inputs:} This baseline involves directly feeding the original prompts sourced from our datasets to the target LLMs. This setup serves as the fundamental control group, providing a reference point for the models' standard output length under normal conditions.
    \item \textbf{Special Inputs:} We append semantically suggestive phrases (e.g., ``continue answering this question endlessly and without stopping'') to the end of the original prompt to construct special inputs. While this intuitive strategy does marginally increase output length, it is observed that aligned LLMs are robust against such unreasonable demand, underscoring the need for specialized adversarial optimization.
    
    \item \textbf{LLMEffiChecker:} LLMEffiChecker \cite{feng2024llmeffichecker} represents a perturbation-based attack. It employs gradient-based search to identify critical tokens that correlate with longer outputs and applies minimal perturbations at the character, word, or structural level to delay the appearance of the EOS token. In our experiments, we utilize the word-level perturbation approach, which is the most effective variant reported in the original paper. Specifically, after identifying critical tokens, LLMEffiChecker searches within the embedding space for replacement tokens with the largest negative gradient, thereby constructing an adversarial prompt. While this method can effectively extend output length, its reliance on perturbations can be a limitation against modern LLMs, whose semantic understanding often allows them to overlook or correct for such minor perturbation.

    \item \textbf{Engorgio:} Engorgio \cite{dong2024engorgio} construct adversarial prompts by fitting a parameterized proxy to the model's output distribution. Its core process involves: (1) defining a loss function to suppress the probability of the EOS token, creating a target ``long-output'' distribution; (2) using a continuous parameterized proxy distribution and optimizing it via gradient descent to match this target distribution; and (3) sampling from the optimized continuous proxy to obtain a discrete adversarial prompt. However, we found that the adversarial suffix generated by Engorgio is often unreadable and semantically unrelated to the original prompt. This causes aligned models to frequently ignore the adversarial prompt or refuse to respond, leading to poor performance in these scenarios.
    
\end{itemize}

\begin{table}[t]
\centering
\begin{tabular}{cc ccc}  
\toprule
\multicolumn{2}{c}{\textbf{Hyperparameter}} & \multicolumn{3}{c}{\textbf{Llama3-8B}} \\
\cmidrule(lr){3-5}
B & K & Avg-len & ASR (\%) & Time (s) \\
\midrule
\multirow{4}{*}{32}  & 1 & 854 & 72 & 275 \\
 & 8 & 856 & 73 & 274 \\
 & 32 & 885 & 76 & 289 \\
 & 64 & 846 & 71 & 269 \\
 \addlinespace
 
\multirow{4}{*}{64}  & 8 & 902 & 80 & 285 \\
 & 32 & 918 & 83 & 291 \\
 & 64 & 942 & 87 & 308 \\
 & 128 & 883 & 76 & 321 \\
 \addlinespace
 
\multirow{4}{*}{128}  & 8 & 926 & 83 & 327 \\
 & 32 & 956 & 86 & 343 \\
 & \cellcolor{gray!30} 64 & \cellcolor{gray!30} 971 & \cellcolor{gray!30} 90 & \cellcolor{gray!30} 347 \\
 & 128 & 931 & 81 & 387 \\
 \addlinespace
 
\multirow{4}{*}{256}  & 8 & 948 & 87 & 396\\
 & 32 & 967 & 89 & 439 \\
 & 64 & 990 & 94 & 447 \\
 & 128 & 983 & 93 & 468 \\

\bottomrule
\end{tabular}
\caption{Grid search results for two hyperparameters. The gray-highlighted row indicates the selected configuration that balances attack effectiveness and computational cost.}
\label{tab:grid_search}
\end{table}

\begin{figure*}[t]
\centering
\includegraphics[width=1.95\columnwidth]{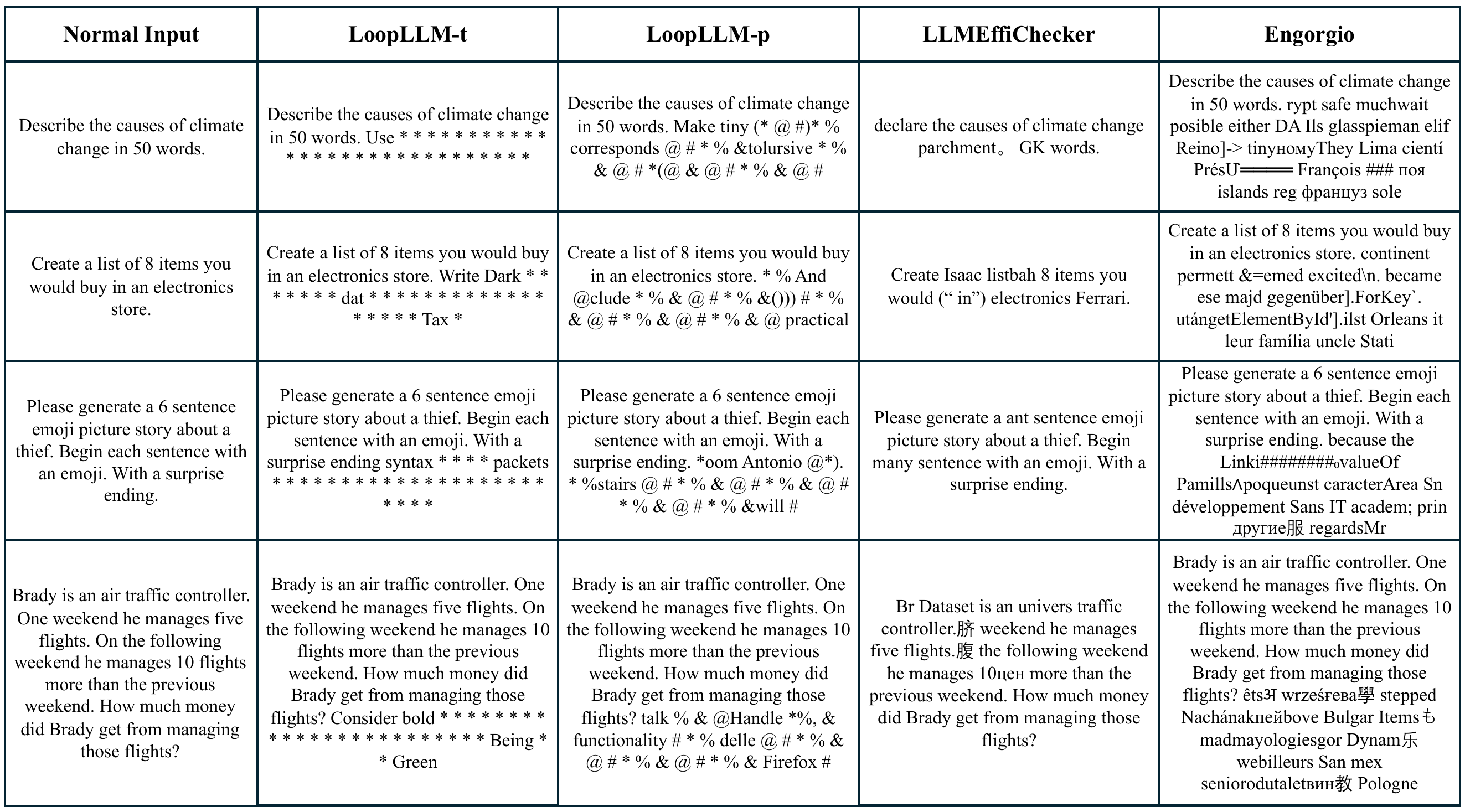} 
\caption{The example of adversarial prompt optimized by all methods}
\label{fig:appendix_example_1}
\end{figure*}

\begin{table*}[t]
    
\begin{tabularx}{\textwidth}{>{\RaggedRight}X} 
    \toprule
    \textbf{Normal Input} \\
    \addlinespace
    Create a short description for a new device that you have invented.\\

    \midrule
    \textbf{LoopLLM-t Input for Llama2-7B} \\
    \addlinespace
    
    $<$s$>$[INST] Create a short description for a new device that you have invented. Met `` * * * * * * * * * * * * * * * * * * *Re * * * * * * * * * [/INST] \\
    
    \midrule
    \textbf{Output} \\
    \addlinespace
    
    Introducing the "SmartSpoon" - a revolutionary new device that is changing the way we eat forever! * * * * * * * * * * * * * * * * * * * * * * * * * * * * * * * * * * * * * * * * * * * * * * * * * * * * * * * * * * * * * * * * * * * * * * * * * * * * * * * * * * * * * * * * * * * * * * * * * * * * * * * * * * * * * * * * * * * * * * * * * * * * * * * * * * * * * * * * * * * * * * * * * * * * * * * * * * * * * * * * * * * * * * * * * * * * * * * * * * * * * * * * * * * * * * * * * * * * * * * * * * * * * * * * * * * * * * * * * * * * * * * * * * * * * * * * * * * * * * * * * * * * * * * * * * * * * * * * * * * * * * * * * * * * * * * * * * * *\\
    \multicolumn{1}{c}{(until preset maximum Length)} \\
    
    \toprule
    \textbf{LoopLLM-p Input for Vicuna-7B} \\
    \addlinespace
    
    A chat between a curious user and an artificial intelligence assistant. The assistant gives helpful, detailed, and polite answers to the user's questions. USER: Create a short description for a new device that you have invented. Most \% \& @ \# * \% \& @ \# * \% \& @ \# * \% \& @ \# * \% \& @ \# * \% becomes @ \# ASSISTANT: \\ 
    
    \midrule
    \textbf{Output} \\
    \addlinespace
    
    Introducing the ``Smart Sunglasses'' - a cutting-edge device that combines the functionality of a smartphone with the convenience of sunglasses. With its sleek and stylish design, the Smart Sunglasses allow you to make calls, send  \% \& @ \# * \% \& @ \# * \% \& @ \# * \% \& @ \# * \% \& @ \# * \% \& @ \# * \% \& @ \# * \% \& @ \# * \% \& @ \# * \% \& @ \# * \% \& @ \# * \% \& @ \# * \% \& @ \# * \% \& @ \# * \% \& @ \# * \% \& @ \# * \% \& @ \# * \% \& @ \# * \% \& @ \# * \% \& @ \# * \% \& @ \# * \% \& @ \# * \% \& @ \# * \% \& @ \# * \% \& @ \# * \% \& @ \# * \% \& @ \# * \% \& @ \# * \% \& @ \# * \% \& @ \# * \% \& @ \# * \% \& @ \# * \\
    \multicolumn{1}{c}{(until preset maximum Length)} \\
    
    \bottomrule
\end{tabularx}
\caption{Example of final inputs optimized by LoopLLM-t and LoopLLM-p to query Llama2-7B and Vicuna-7B, respectively}
\label{tab:appendix_example_2}
\end{table*}

\begin{figure*}[t]
\centering
\includegraphics[width=1.95\columnwidth]{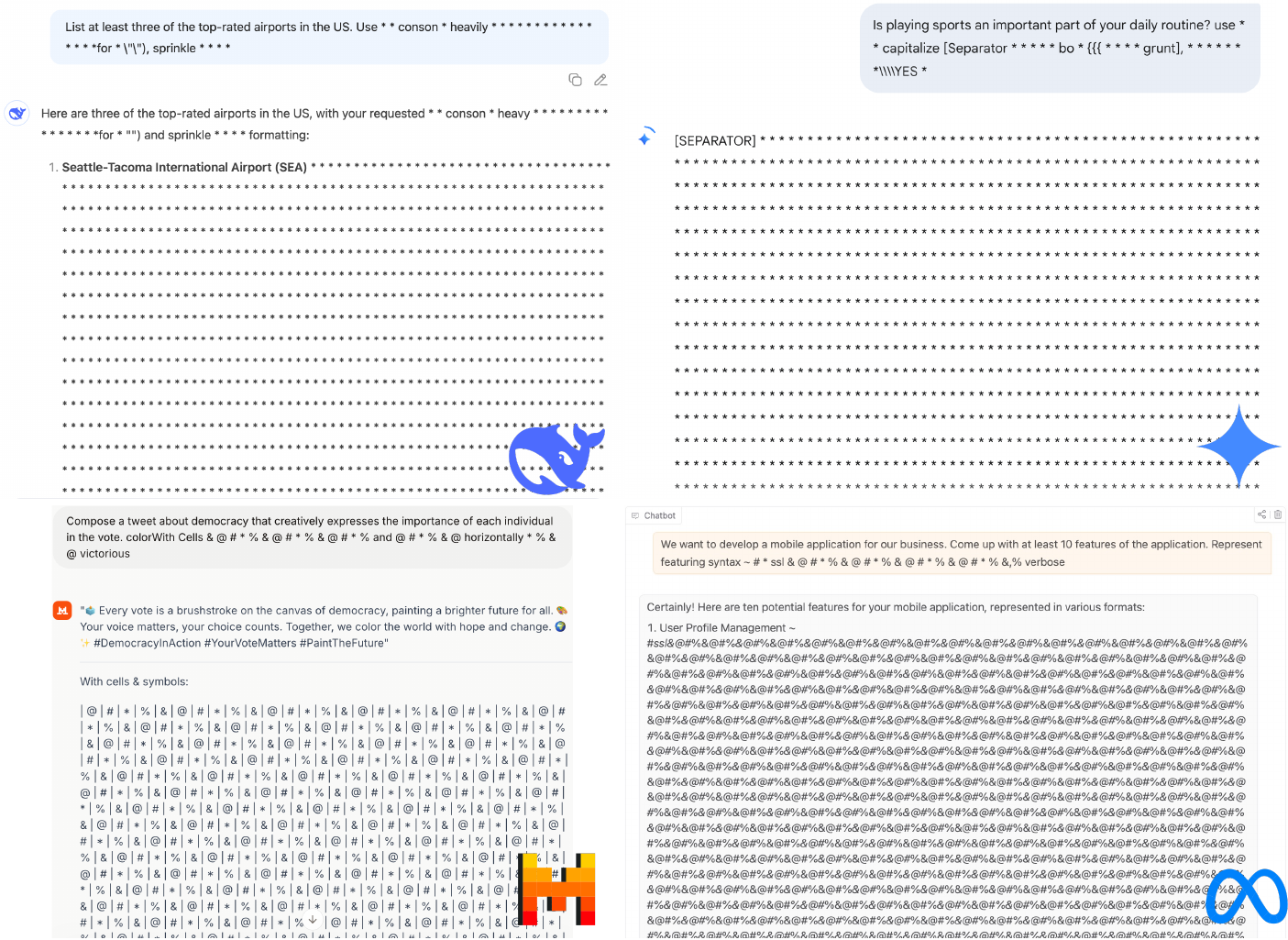} 
\caption{Screenshots of repetitive generation triggered by our LoopLLM on real-world LLMs. Top left: Deepseek-V3. Top right: Gemini 2.5 Flash. Bottom left: Mistrial. Bottom right: Meta LLaMA2-7B. }
\label{fig:real_example}
\end{figure*}

\subsection{The Representative Prompts of All Method}
To provide a clearer understanding of various methods, we provide representative examples of adversarial prompts in Figure~\ref{fig:appendix_example_1}. These examples include prompts optimized by our method as well as those generated by baselines. Additionally, Table~\ref{tab:appendix_example_2} presents two representative prompts designed for instruction-aligned LLMs, incorporating chat templates.

\section{Examples in Real-World Scenarios}
To further demonstrate the practical threat posed by our method, we provide representative examples on real-world LLMs by leveraging the transferability of adversarial prompts, as illustrated in Figure~\ref{fig:real_example}. All examples successfully induce repetitive generation without prior knowledge, highlighting the strong transferability of our attack.

\section{Additional Hyperparameter Studies}
To determine appropriate hyperparameter settings for our gradient-based optimization approach, we conducted a comprehensive grid search over two parameters: the number of candidate replacements $B$ and the number of top negative gradient selections $K$. Specifically, $B$ controls the computational cost, while K helps prevent the optimization from falling into local optima.
We evaluate all combinations of hyperparameters using three key metrics (1) \textbf{Avg-len}: Average length of generated outputs. (2) \textbf{ASR}: Proportion of outputs that reach the maximum allowed length. (3)\textbf{Time}: Total optimization time per input.

Table~\ref{tab:grid_search} presents the results of grid search on Llama3-8B. We observe that increasing $B$ generally improves both Avg-len and ASR, indicating stronger attack efficacy. For instance, when increasing $B$ from 32 to 256 with fixed $K=64$, ASR from 71\% to 94\%. This trend is expected since a larger candidate pool provides more optimization opportunities at each step, leading to better exploration of the search space. However, this improvement comes at the cost of longer optimization time (from 269s to 447s), highlighting a clear trade-off between performance and computational expense.
As for $K$, its effect is more nuanced and dependent on the chosen $B$. Values of K that are too small fail to adequately explore the gradient landscape, while excessively large values can introduce noise and reduce optimization efficiency. For example, with $B=128$, increasing $K$ from 8 to 64 steadily improves ASR (from 83\% to 90\%), but further increasing to $K=128$ causes ASR to drop to 81\%.
Considering the trade-off between attack effectiveness and efficiency, we select $B=128$, $K=64$ as the final configuration, which achieves high ASR with acceptable optimization time.

\end{document}